\documentclass[a4paper]{article}
\usepackage{float}
\usepackage{Odyssey2022}
\usepackage{epsfig,amssymb,amsmath}
\ninept

\title{The SpeakIn System Description for CNSRC2022}

\name{Yu Zheng, Yihao Chen, Jinghan Peng, Yajun Zhang, Min Liu, Minqiang Xu\textsuperscript{\dag}}

\address{SpeakIn Technologies Co. Ltd.  \\
{\small \tt \{zhengyu, chenyihao, pengjinghan, zhangyajun, liumin, xuminqiang\}@speakin.ai}}%
\begin{document}
\maketitle

\begin{abstract}
This report describes our speaker verification systems for the tasks of the CN-Celeb Speaker Recognition Challenge 2022 (CNSRC 2022). This challenge includes two tasks, namely speaker verification(SV) and speaker retrieval(SR). The SV task involves two tracks: \emph{fixed track} and \emph{open track}. In the fixed track, we only used \emph{CN-Celeb.T} as the training set. For the open track of the SV task and SR task, we added our open-source audio data. The ResNet-based, RepVGG-based, and TDNN-based architectures were developed for this challenge. Global statistic pooling structure and MQMHA pooling structure were used to aggregate the frame-level features across time to obtain utterance-level representation. We adopted AM-Softmax and AAM-Softmax combined with the Sub-Center method to classify the resulting embeddings. We also used the Large-Margin Fine-Tuning strategy to further improve the model performance. In the backend, Sub-Mean and AS-Norm were used. In the SV task fixed track, our system was a fusion of five models, and two models were fused in the SV task open track. And we used a single system in the SR task. Our approach leads to superior performance and comes the 1st place in the open track of the SV task, the 2nd place in the fixed track of the SV task, and the 3rd place in the SR task.


\end{abstract}
\section{Data}
\subsection{Training dataset}
\label{ssec:subhead}
For Task 1 SV \emph{fixed track}, only the \emph{CN-Celeb.T} was used to perform system development. We concatenate the utterances from the same speaker. We filtered for speakers whose concatenated audio duration is too short and obtained 2745 speakers. We split the concatenated utterance into short segments by 12 seconds. We here adopted a 3-fold speed augmentation at first to generate extra twice speakers. Each speech segment in this dataset was perturbed by 0.9 or 1.1 factor based on the SoX speed function. Then we obtained 8,235 speakers, which is the triple amount of the original speakers.

For Task 1 SV \emph{open track} and Task 2 SR \emph{open track}, we use CN-Celeb.T dataset with speed perturbation, VoxCeleb dataset with speed perturbation, and our own open-source Chinese audio data. In total, there are 245,497 speakers in this dataset. The datasets used for training included:
\begin{enumerate}
    \item CN-Celeb.T: contains 797 speakers from CN-Celeb1 dev \cite{cnceleb1} and 1,996 speaker from CN-Celeb2 \cite{cnceleb2}. The utterances are from 11 genres, namely: advertisement, drama, entertainment, interview, live broadcast, movie, play, recitation, singing, speech, and video blog.
    \item VoxCeleb: contains 5,994 speakers from VoxCeleb-2\cite{chung2018voxceleb2}.
    \item Our own open-source audio data.
\end{enumerate}

We applied the following techniques to augment each utterance:
\begin{itemize}
    \item Reverberation: artificially reverberation using a convolution with simulated RIRs\cite{rirs} from the AIR dataset.
    \item Music: taking a music file (without vocals) randomly selected from MUSAN\cite{snyder2015musan}, trimmed or repeated as necessary to match duration, and added to the original signal (5-15dB SNR).
    \item Noise: MUSAN noises were added at one second intervals throughout the recording (0-15dB SNR).
    \item Babble: MUSAN speech was added to the original signal (13-20dB SNR).
\end{itemize}

\subsection{Evaluation dataset}
The evaluation set of SV task, including fixed track and open track, is CN-Celeb.E dataset, which is a subset of 200 speakers from CN-Celeb1. The test utterance is a single utterance from one speaker. The enroll utterance may be a utterance from one speaker or the utterance concatenated by multiple utterance from the same speaker. 

For the SR task, the SR.dev and SR.eval are released. SR.dev consists of 5 target speakers, each with 10 test utterances. SR.eval consists of 25 target speakers, each with 10 test utterance. The number of non-target utterances is 20,000 in SR.dev and 500,000 in SR.eval. SR.eval is released for system evaluation.

\subsection{Feature}
\label{ssec:subhead}
We extracted both 81-dimensional log Mel filter bank with energy based on Kaldi. The window size is 25 ms, and the frame-shift is 10 ms. 200 frames of features were extracted without extra Voice Activation Detection (VAD). All features were cepstral mean normalized (CMN) in both our training modes.


%
\section{Models}

\subsection{Backbone}
\subsubsection{ResNet}
\label{ssec:subhead}
As one of the most classical ConvNets, ResNet \cite{he2016deep} has proved its power in speaker verification. In our systems,  bottleneck-block-based ResNet (deeper structures: ResNet-74, ResNet-101) are adopted. Base channels of all these ResNets are 64.

\subsubsection{RepVGG}
\label{ssec:subhead}
In our previous work, we have proved that the RepVGG structure \cite{zhao2021speakin} is very effective in speaker recognition. We select RepVGG-A2 as our backbones. The model adopts 64 base channels.

\subsubsection{ECAPA-TDNN}
\label{ssec:subhead}
The ECAPA-TDNN\cite{thienpondt2020idlab} structure is a modified Time Delay Neural Network structure and has been shown to perform well in speaker recognition. We use 2048 base channels in this competition.

\subsection{Pooling method}
\label{ssec:subhead}
The pooling layer aims to aggregate the variable sequence to an utterance level embedding. In addition to global statistics pooling layer (GSP), was also used multi-query multi-head attention pooling mechanism layer (MQMHA) \cite{zhao2021speakin} to aggregate along the time across. In our systems, RepVGG-A2 is followed by GSP, and other backbones are followed by MQMHA.

\subsection{Loss function}
\label{sssec:subhead}

Recently, margin based softmax methods have been widely used in speaker recognition works. To make a much better performance, we strengthen the AM-Softmax \cite{wang2018additive, wang2018cosface} and AAM-Softmax \cite{deng2019arcface} loss functions by Sub-Center method.

The Sub-Center method \cite{deng2020sub} was introduced to reduce the influence of possible noisy samples. The formulation is given by:

\begin{equation}
  cos(\theta_{i,j})=\max_{1\leq k\leq K}(||x_i||\cdot||W_{j,k}||)
  \label{eq3}
\end{equation}
where the $\max$ function means that the nearest center is selected and it inhibits possible noisy samples interfering the dominant class center, $K$ is the number of sub-centers of each class.

\section{Training Protocol}
We used Pytorch \cite{paszke2019pytorch} to conduct our experiments. All of our models were trained through two stages. 

In the first stage, the SGD optimizer with a momentum of 0.9 and weight decay of 1e-3 (4e-4 for online training) was used. We used 8 GPUs with 1,024 mini-batch and an initial learning rate of 0.08 to train all of our models. 200 frames of each sample in one batch were adopted. We adopted ReduceLROnPlateau scheduler with a frequency of validating every 2,000 iterations, and the patience is 2. The minimum learning rate is 1.0e-6, and the decay factor is 0.1. Furthermore, the margin gradually increases from 0 to 0.2 \cite{liu2019large}. We used Sub-Center in the loss function in this stage. The $K$ of the Sub-Center is set to 3.

In the Large-Margin Fine-Tuning stage (LM-FT) \cite{thienpondt2020idlab}, settings are slightly different from the first stage. Firstly, we removed the speed augmented part from the training set to avoid domain mismatch. Secondly, we changed the frame size from 200 to 1200 and increased the margin exponentially from 0.2 to 0.8. The AM-Softmax loss was replaced by AAM-Softmax loss. Thirdly, we choose the center with the largest norm as the "dominated center" and discard the other sub-centers. We only use the selected dominated center as the speaker center throughout the whole fine-tuning process. Finally, we adopted a smaller fine-tuning learning rate of 8e-5 and a 256 batch size. The learning rate scheduler is almost the same while the decay factor became 0.5.

For the SV task open track and SR task, we used our own open-source data and VoxCeleb data as training datasets in the first stage. Then we added CN-Celeb.T data in the fine-tuning stage. We directly added extra weights of speaker centers for CN-Celeb.T data in the last classification layer of the model and randomly initialed them. All weights of the whole model are fine-tuned in the LM-FT stage.



\section{Backend}
We used cosine distance for scoring in both tasks. In addition, Sub-Mean method and adaptive symmetric score normalization (AS-norm)\cite{asnorm} were used in both tasks.

\subsection{Sub-Mean }
The Sub-Mean method was used for all our models. We extracted the embedding vectors from the training data to compute the mean embedding vector of each genre. The enrollment and test embedding vectors both subtract the mean embedding vector corresponding to the same genre before scoring:
\begin{equation}
s(x_{e} ,x_{t}) = cos(x_{e}-\bar{x}_i ,x_{t}-\bar{x}_j)
\end{equation}

where $x_{e}$, $x_{t}$ are enrollment and test embedding vectors respectively, $s(x_{e}, x_{t})$ is the similarity score between $x_{e}$ and $x_{t}$, and $\bar{x}_i$, $\bar{x}_j$ are the genre mean embedding vector corresponding to the same genre of enrollment and test utterances respectively. For the enrollment utterance that is concatenated by different genre segments, we choose the genre with the largest number of segment as its genre.

\begin{table*}[t]
\centering
\caption{ Performance on CN-Celeb.E set in SV task fixed track.}
\label{tab:tabel_two}
\setlength{\tabcolsep}{10.5mm}{
\resizebox{\textwidth}{!}{
\begin{tabular}{ccccc}
\hline
System      & \multicolumn{2}{c}{Cosine Similarity} & \multicolumn{2}{c}{Sub-Mean\&AS-Norm} \\
            & EER               & minDCF            & EER               & minDCF            \\ \hline
ECAPA-TDNN  & 8.8257            & 0.4086            & 8.6623            & 0.3914            \\
RepVGG-A2   & 7.7387            & 0.3681            & 7.4233            & 0.3460            \\
ResNet34    & 7.8231            & 0.3755            & 7.4571            & 0.3518            \\
ResNet74    & 7.7274            & 0.3785            & 7.3162            & 0.3475            \\
ResNet101   & 6.2518            & 0.3619            & 6.1335            & 0.3358            \\ \hline
Fusion       & \textbf{-}        & \textbf{-}        & \textbf{5.9530}   & \textbf{0.3185}   \\ \hline
\end{tabular}
}
}
\end{table*}

\begin{table*}[t]
\centering
\caption{ Performance on CN-Celeb.E set in SV task open track.}
\label{tab:tabel_two}
\setlength{\tabcolsep}{10.5mm}{
\resizebox{\textwidth}{!}{
\begin{tabular}{ccccc}
\hline
System    & \multicolumn{2}{c}{Cosine Similarity} & \multicolumn{2}{c}{Sub-Mean\&AS-Norm} \\
          & EER               & minDCF            & EER               & minDCF            \\ \hline
RepVGG-A2 & 5.8519            & 0.2925            & 6.0321            & 0.2711            \\
ResNet101 & 5.0577            & 0.2574            & 5.0183            & 0.2418            \\ \hline
Fusion    & \textbf{-}                 & \textbf{-}                 & \textbf{4.6630}            & \textbf{0.2384}            \\ \hline
\end{tabular}
}
}
\end{table*}

\subsection{AS-Norm }
AS-Norm was used for all systems. For AS-Norm, we selected the original CN-Celeb.T dataset without any augmentation. The cohort was created by using the speaker's random one utterance embedding vector as a speaker center and consisted of 2745 speaker centers. Only part of the cohorts are selected to compute mean and standard deviation for normalization, and top-300 highest scores are selected in our systems.

\subsection{Fusion }
Fusion was performed by computing the weighted average of the score of each individual system. The fusion score is used as our final submission for both track of SV task.

\begin{table}[]
\centering
\caption{ Performance on SR.eval set.}
\label{tab:tabel_two}
\setlength{\tabcolsep}{13.5mm}{
\begin{tabular}{cc}
\hline
System    & mAP    \\ \hline
RepVGG-A2 & 0.5203 \\
ResNet101 & \textbf{0.6106} \\ \hline
\end{tabular}
}
\end{table}

\begin{table}[]
\centering
\caption{ Performance of RepVGG-A2 model with different Sub-Center methods in both training stages in SV task.}
\label{tab:tabel_two}
\setlength{\tabcolsep}{4.4mm}{
\begin{tabular}{clcc}
\hline
Stage      & \multicolumn{1}{c}{Sub-Center} & \multicolumn{2}{c}{Cosine Similarity} \\
           & \multicolumn{1}{c}{}           & EER               & minDCF            \\ \hline
Baseline   & k=3                            & 8.8313            & 0.4226            \\
LM-FT & k=3                            & 8.2625            & 0.3810            \\
LM-FT & k=1(MAX)                       & \textbf{7.7387}            & \textbf{0.3780}            \\
LM-FT & k=1(SUM)                       & 7.5472            & 0.3807            \\ \hline
\end{tabular}
}
\end{table}

\section{Results}
\subsection{SV task}
Results of experiments on all our systems developed for the fixed track and open track of the SV task are displayed in Table 1 and Table 2 respectively. The performance is measured on the CN-Celeb.E set in terms of Equal Error Rate (EER) and Minimum Detection Cost(minDCF) with a prior target probability, $P_{tar}$ of 0.01. For SV task, all our experiments are based on raw enroll and test utterances from CN-Celeb.E set provided by the CNSRC2022 organizer.

For the fixed track, we trained five systems. The results of all considered single systems are presented in Table 1. Our best single system is ResNet101, which has a 6.13\% EER and 0.335 minDCF after Sub-Mean and AS-Norm calibration. ECAPA-TDNN is the worst of all single models but it is helpful for model fusion. The final fused result used all five single systems and get a 5.95\% EER and 0.3185 minDCF. We ranked second in the fixed track.

For the open track, we only trained two individual systems (see Table 2). The best single system result is ResNet101, which has a 5.02\% EER and 0.2418 minDCF. The fused result of both systems is 4.66\% EER and 0.2384 minDCF. From the results in Table 1 and Table 2, the combined use of the Sub-Mean method and the AS-Norm method can relatively reduce the minDCF value by 4.2\%-8.1\%. Finally, We ranked first in the open track.

In the first stage, we use the Sub-Center method, in which the number of sub-centers is set to 3. This system has a 0.4226 minDCF. For the LM-FT stage, we experimented with three Sub-Center methods (Table 4). In the first method, we also used the Sub-Center method to fine-tune the model in the LM-FT stage. This system has a 0.3810 minDCF. The number of sub-centers remains the same as in the first stage. In the second method, we select the sub-center with the largest norm among all the sub-centers of the same speaker as the only speaker center and discarded the other sub-centers. This system has a 0.3780 minDCF. In the third method, we sum all sub-center of the same speaker and used the summed center vector as the new speaker center. The new speaker center as the only center of the speaker is tuned in the LM-FT stage. This system has a 0.3807 minDCF. The experimental results from Table 4 show that the three methods have improved the model performance, of which the second method is the best.

\subsection{SR task}
The experimental results of our SR systems are described in Table3. We developed two single systems for the SR task. RepVGG-A2 get a 52.03\% mAP and ResNet101 has a 61.06\% mAP. As a result, we used the ResNet101 system as our final submission for the SR task and ranked third in the SR task.


%


%
\section{Conclusion}
We experimented multiple models on the SV and SR tasks, and ResNet network achieved the best results on both tasks. 

We increased the number of utterances and the number of speakers through data augmentation. We changed hyper-parameters and training data in the LM-FT stage. All these strategies improve the performance of the model. We use Sub-Centers in the first stage, and select the sub-center with the largest norm among the sub-centers of the same speakers as the only speaker center in the LM-FT stage. This strategy further improves the performance of the model. And this method allows us to use larger chunk-size and margin values during the LM-FT stage.

In the backend, We used the Sub-Mean method and AS-Norm method for score calibration. Then we used a fusion strategy by simply weighted averaging the scores of multiple systems. The performance of the fused system is better than that of all single systems. These backend methods all further reduce the minDCF value on CN-Celeb.E set. 

%

\bibliographystyle{IEEEbib}
\bibliography{Odyssey2022_BibEntries}


\end{document}